\documentclass[12pt]{article}

\usepackage{amsmath,amssymb,titling,authblk}
\usepackage{amsmath}
\usepackage{amscd}
\usepackage{cite}

\usepackage{graphicx}

\usepackage[normalem]{ulem}

\newcommand{\be}{\begin{equation}}
\newcommand{\ee}{\end{equation}}
\newcommand{\bea}{\begin{eqnarray}}
\newcommand{\eea}{\end{eqnarray}}

\begin{document}




\pretitle{\vskip -100pt\
\begin{flushright}\small
ICCUB-15-017
\end{flushright}\vspace*{2pc}%
\begin{center}\LARGE}

\title{Inconstant Planck's constant}


\author[1]{Gianpiero Mangano$^*$}
\author[1,2,3]{Fedele Lizzi}
\author[4]{Alberto Porzio}
\affil[1]{INFN, Sezione di Napoli, Complesso Univ.\ Monte S. Angelo, Via Cintia, I-80126 Napoli, Italy}
\affil[2]{Dipartimento di Fisica, Universit\`a di Napoli ``Federico II'', Complesso Univ.\ Monte S. Angelo, Via Cintia, I-80126 Napoli, Italy}
\affil[3]{Departament de Estructura i Constituents de la Mat\`eria, Institut de Ci\'encies del Cosmos, Univ.\ de Barcelona, Spain}
\affil[4]{ CNR-SPIN, Unit\`a di Napoli, Complesso Univ.\ Monte S.~Angelo, Via Cintia, I-80126 Napoli, Italy}
\date{}
\maketitle
 \let\thefootnote\relax\footnote{mangano@na.infn.it}
 \let\thefootnote\relax\footnote{fedele.lizzi@na.infn.it}
 \let\thefootnote\relax\footnote{porzio@na.infn.it}
\begin{abstract}
Motivated by the Dirac idea that fundamental constant are dynamical variables and by conjectures on quantum structure of spacetime at small distances, we consider the possibility that Planck constant $\hbar$ is a time depending quantity,  undergoing random gaussian fluctuations around its measured constant mean value, with variance $\sigma^2$ and a typical correlation timescale $\Delta t$. We consider the case of propagation of a free particle and a one--dimensional harmonic oscillator coherent state, and show that the time evolution in both cases is different from the standard behaviour. Finally, we discuss how interferometric experiments or exploiting coherent electromagnetic fields in a cavity may put effective bounds on the value of $\tau= \sigma^2 \Delta t$.
\end{abstract}






\section{Introduction} \label {sec:intro}
The idea that  fundamental constants in physics may be spacetime varying parameters  was first suggested by Dirac  with his ``Large Numbers Hypothesis''~\cite{Dirac}. Since then, many efforts have been devoted to embed this idea in coherent theoretical frameworks. One possibility is that these constants may be \emph{effective} quantities. This happens, for example, in higher-dimensional theories, after reduction of coupling constants to the four-dimensional subspace. For a critical summary of this subject see~\cite{Uzan:2002vq, Uzan:2010pm}. 

The most studied examples of varying fundamental constant are perhaps the Newton gravitational constant $G_N$ and the fine structure coupling $\alpha$. In the latter case there is quite a large number of experimental constraints on its time evolution, spanning several order of magnitude in redshift, or cosmological time, from laboratory experiments, geophysical tests as the Oklo phenomenon,  the anisotropies of Cosmic Microwave Background, till the early epoch of Big Bang Nucleosynthesis, see~\cite{Uzan:2010pm} and references therein. 
Spatial and time variations of $\alpha$  have been also constrained by observations of absorption spectral lines in intervening clouds along the line of sight of  distant Quasars at redshifts $z \sim 2 -3$, ~\cite{Webb:2000mn,Murphy:2003hw,Webb:2010hc,King:2012id,Wilczynska:2015una}. 
 
Evidences for  dynamical features of fundamental constants would have of course, an enormous impact on the general picture of physics laws. The typical approach to this theoretical and experimental issue, as in the case of $\alpha$ just mentioned, is to look for signals of variation over long time period, billion years in the case one uses cosmological observables. The underlying idea is that the dynamical field which describes a particular fundamental constant is changing as a {\it classical} background field under the action of some effective potential which drives the field towards its minima. However, if this putative field $\Psi$ reaches a stable configuration in very short time intervals, so that there is no further evolution in the long time regime, we could only see an effect by strongly perturbing this configuration and moving $\Psi$ away from the minimum. This can be achieved by exploiting the $\Psi$ coupling with matter fields and exciting the field by pumping energy into it, i.e. the typical accelerator approach to unveil new degrees of freedom. However, even if $\Psi$ seats at the classical potential minimum, we expect its value to undergo small random fluctuations. In this case the question is how these fluctuations might produce observable effects in experiments. 

We recall that many theoretical arguments point out that at short time and distance scales both quantum field theory and gravitational interactions described by general relativity ought to be embedded in some form of \emph{quantum gravity}, a theory not yet known. Whatever is this theory, it will deeply change the geometry of spacetime and phase space. Examples are  theories in which spacetime is composed by branes, feature a noncommutative geometry or are characterized by the spin foam of loop quantum gravity, see, e.g. ~\cite{Doplicher:1994tu,Seiberg:1999vs} and~\cite{Rovelli:2010bf} and references therein. The role of a spacetime fuzziness and stochastic Lorentz invariance violation has been recently studied in~\cite{Vasileiou:2015wja}.

Since its introduction at the dawn of last century \cite{Planck1901}, $h$ has been considered one of the fundamental \textit{constant} ruling physical phenomena.
Its value has been measured by different methods spanning the spectrum of black body radiation, kinetic energy of photoelectrons in photoelectric effect, and more recently, X-ray density crystal measurement, and watt balance. The latter two have recently achieved a relative precision of $1.2$ parts in $10^{8}$ \cite{Shi-Song2015}. The actual accepted value is $h=6.626 070 040(81) \cdot 10^{-34}$ J s \cite{codata}.
The precise measurement of $h$ plays a crucial role in establishing both a new SI and a revised fundamental physical constant system.

The (reduced) Planck constant $\hbar$ enters into the definition of the Heisenberg uncertainty principle, one of the pillars of the probabilistic nature of quantum mechanics. In fact, deformations of the quantum phase space have been put forward in the form of generalized uncertainty principle,  see e.g.~\cite{Amati:1988tn, Konishi:1989wk,Maggiore:1993rv,Kempf:1994su}, described by the commutation relation
\be
[\mathbf x^i,\mathbf p_j]=i\,\hbar\,\left(\delta^i_j+ f^i_j(\mathbf x,\mathbf p)\right) \, , \label{xpcommold}
\ee
where $f^i_j$ depend upon some ``small parameter''. This is particular case of a most general phase space commutation relation among position and momenta which also includes
\begin{eqnarray}
[\mathbf x^i,\mathbf x^j]&=&i\theta^{ij}\nonumber\\
{}[\mathbf p_i,\mathbf p_j]&=&iC_{ij} \label{xxppcomm}
\end{eqnarray}
where $f,\theta~C$ are such that Jacobi identities are satisfied.
In general, these relations break Lorentz invariance, and a way to restore it  to consider $f,\theta$ and $C$ to be \emph{random} variables with microscopic variations, so that for large distances and time intervals the invariance is restored. The commutations relations~\eqref{xpcommold} and~\eqref{xxppcomm} mean that the phase space has a non canonical symplectic structure, but with a Darboux transformation it is always possible to express (locally) the structure as a nontrivial commutator between position and momenta, all other commuters being zero. Since the transformation mixes positions with momenta this calls into question the meaning of the position and momentum. We will consider that \emph{at any given instant} the observables can be seen as generated by two sets of operators, $\bf x^i$ and $\bf p_j$, with commutation relations as in (\ref{xpcommold}).

If one considers that corrections to the commutators are small and  depend on the background and fluctuations at some high energy scale, the net result is that the r.h.s. of commutation relations can be effectively described by a  random function acting as an effective dynamical field. This is tantamount to consider $\hbar$ as a fluctuating parameter. In this paper we will consider that the space fluctuations are integrated out, and will concentrate on the time variations. Energy  scale dependence of $\hbar$ induced by nontrivial commutation relations has been proposed in~\cite{Calmet}. In~\cite{Hossenfelder} the possibility that $\hbar$ is a field whose expectation value can vanish at high temperatures was considered as a possibility to quantize gravity.
 
A comment about the issue of dimensionality. Planck constant is not adimensional, and claims about its time dependence would be always related to some particular combinations of $\hbar$ and other dimensionful quantities, depending on the adopted experimental method. Only for adimensional combinations of fundamental constants, such as $\alpha$, it is meaningful to speak about spacetime variation unambiguously. The variability of fundamental constants is a delicate issue, see e.g.~\cite{Duff}. We will see next that the stochastic nature of $\hbar$ is encoded in a {\it dimensionful}  time parameter $\tau$ and observable effects for light propagation or a harmonic oscillator show up  through the dimensionless combination $\omega \tau$, with $\omega$ the light or oscillator frequency.

Our aim in this paper is not to develop a specific theoretical framework describing this scenario. We will only assume that there is no classical background evolution, but the field nature of the Planck constant shows up in random fluctuation around the constant measured mean value, which we will continue to denote by $\hbar$.

\section{General formalism}

Random fluctuations of $\hbar$ are introduced via an adimensional gaussian stochastic variable $\varepsilon(t)$, so that the {\it effective} Planck constant reads $\hbar (1+\varepsilon(t))$, with 
\bea 
\overline{\varepsilon(t)} &=&0 \, , \label{mean0} \\
\overline{\varepsilon(t) \varepsilon(t')} &=& \tau  \, \delta(t-t')\, . \label{uncorr}
\label{epsprop}
\eea
Overline denotes the mean over $\varepsilon$  probability distribution. The parameters are as follow: $\tau=\sigma^2 \Delta t$, $\sigma$ is the variance and equation (\ref{uncorr}) states that fluctuations are uncorrelated for time differences larger than a typical correlation time $\Delta t$, analogous of  the strength of fluctuation force in brownian motion dynamics in Stokes--Einstein theory for diffusion coefficient.
These relations are time translation invariants. 

Planck constant fixes the commutator of canonically conjugated variables,  position and momentum for non relativistic particles, and bosonic (fermionic) field and conjugate momentum field commutator (anticommutator). 
Consider a particle in one dimension. If $\hbar$ {\it fluctuates}, we have 
\be \left[{\bf x},{\bf p} \right] = i \hbar (1+ \varepsilon(t)) \label{xpcomm} \, .\ee
The commutator with a Hamiltonian ${\bf H}$ gives the time evolution of operators ${\bf A}$ 
\be
\frac{d {\bf A }}{d t}=  \frac{1}{i \hbar} [{\bf A},{\bf H}] + \frac{\partial {\bf A }}{\partial t} \, .\label{newdadt}
\ee
The first term on the r.h.s. now explicitly depends on time via $\varepsilon(t)$, and the time evolution of $\overline{\bf A }$, the average over the probability distribution of $\varepsilon$, shows non--standard behaviour, as we will see. 
Had we defined the Poisson structure by normalizing the commutator to $\hbar(1+ \varepsilon(t))$, all effects cancel and we would be back to standard quantum mechanics. More interesting is the case of equation (\ref{newdadt}), i.e.\ the fundamental Poisson bracket fluctuates. We investigate this possibility.

Our approach is different from {\it stochastic quantization} \cite{PW}, where  a system is considered in thermal contact with some external reservoir, and its quantum properties are obtained in the equilibrium limit. This is achieved by introducing an auxiliary {\it fictious}  time. Equilibrium is reached in the large fictious time limit. In our case, Planck constant depends upon {\it physical} time, leading to novel features.

Quantum measurements is itself a probabilistic concept, and it only makes sense to compute observable averages over the probability distribution given by a wavefunction. We denote this averaging by  $\langle {\bf A} \rangle$. The relation between standard quantum mechanical average and average over $\varepsilon(t)$ distribution is discussed later. 

Commutation relation of equation (\ref{xpcomm}) is represented by defining the action of position and momentum on wavefunctions $\psi(x)$
\bea
 {\bf x} \,\psi(x) &=& A(t) \, x \, \psi(x) = A(t)\, {\bf x}_0\, \psi(x) \, ,\\
 {\bf p} \,\psi(x) &=& -i \hbar \,B(t) \,\frac{d}{dx} \,\psi(x) = B(t)\, {\bf p}_0 \,\psi(x)\, ,
\eea
with $A(t)B(t)=1+\varepsilon(t)$, and ${\bf x}_0$ and ${\bf p}_0$ the canonical pair of standard quantum mechanics.
A change of representation $A(t) \rightarrow A'(t)$ is a time dependent dilation induced by the unitary operator 
\be
U(t) = \exp \left[ \frac{i}{2 \hbar} \log \frac{A'(t)}{A(t)} ({\bf x}_0 {\bf p}_0+{\bf p}_0 {\bf x}_0 ) \right] \, .
\label{unit}
\ee
Here we adopt the choice
\be
A(t) = B(t) =\sqrt{1+ \varepsilon(t)} \, ,
\label{particular}
\ee
and assume that {\it in this representation} equations of motion are given by~(\ref{newdadt}). The rationale is that  effects of any high energy scale theory producing a fluctuating $\hbar$ are expected to disturb phase-space elementary volumes. It seems unnatural that coordinates and momenta should be treated differently, and rescaled by different factors.

Although in principle, all values of $\varepsilon(t)$ are possible, we expect $\varepsilon(t)$ probability distribution to be narrow, so the number of events corresponding to large values is exceedingly small. Since $\varepsilon(t)$ is constant in the time interval $\Delta t$, in a time T its value is randomly extracted $T/\Delta t$ times. For T$ \sim 13.8$ Gyr, the age of the universe, and $\Delta t \geq 5.4 \cdot 10^{-44}$ s, the Planck time, for a gaussian distributed $\varepsilon$, the event number for $\varepsilon < -1$, i.e. imaginary $A(t)$ 
\be
\frac{T}{\Delta t} \, \int_{-\infty}^{-1} \frac{d \varepsilon}{\sqrt{2 \pi \sigma^2}} e^{-\frac{\varepsilon^2}{2 \sigma^2}} = \frac{T}{2 \Delta t} \mbox{erfc}\left(\frac{1}{\sqrt{2 \sigma^2} }\right) \, ,
\ee
is smaller than unity provided  $\sigma \leq 0.06$, a value which seems already too large to be acceptable. It is unlikely that in the whole history of the universe these \emph{large} fluctuations  were ever produced, and we can safely assume $A(t) \in \mathbb{R}$.
 
\section{Experimental strategies: two examples}
\label{examples}
If Planck constant is {\it randomly inconstant}, the issue is to find experimental approaches to constrain the $\tau$, or to find signature of its dynamics. We discuss two possibilities, interferometric experiments and coherent electromagnetic modes in a cavity.

\subsection{Free particles and long baseline interferometric experiments}
\label{free}
Consider a free particle with mass $m$. Schr\"odinger equation reads
\be
{i}{\hbar}  \frac{\partial}{\partial t}\psi =\frac{1}{2 m } (1+\varepsilon(t))\, {\bf p}_0^2 \, \psi \, .\label{schrfree}
\ee
Since $[{\bf p}_0,{\bf H}]=0$, fundamental solutions are still plane waves $\psi_{p_0}$, eigenfunctions of ${\bf p}_0$ with eigenvalue $p_0$. In the representation $L^2(\mathbb{R})$, with $\mathbb{R}$ the ${\bf x}_0$ spectrum
\be
\psi_{p_0}(x,t) = \frac{1}{\sqrt{2 \pi}} \exp \left[i \frac{p_0 x}{\hbar}-i \frac{p_0^2}{2 m \hbar}\left( t + \int_0^t \varepsilon(t') dt' \right) \right] \, .
\label{wf}
\ee
Using these solutions we can construct wave packets. Position and momentum operators act as follows
\bea
&&{\bf x}\,\psi(x,t)=\sqrt{1+\varepsilon(t)} \,x \,\psi(x,t) \, ,\\
&&{\bf p}\,\psi(x,t)= -i \hbar \,\sqrt{1+\varepsilon(t)}\, \frac{d}{d x}\,\psi(x,t) \, .
\eea
We want to evaluate position mean value and uncertainty versus time. 
As mentioned already, averaging should be performed both in the usual quantum sense, and over the $\varepsilon(t)$ probability distribution.
We denote this double average of some observable ${\bf A }$ in the state $|\psi \rangle$ by $\overline{\langle {\bf A } \rangle}_\psi$. For $\Delta t$ much smaller than experimental time resolution, standard quantum mechanics average is already an average over $\varepsilon$ values. Indeed, repeating the measurement on the same state, which defines a {\it quantum} measurement, we also sample $\varepsilon$ probability  distribution. Actually, in quantum mechanics averaging over repeated experiments is the same as an average over simultaneous experiments on a large number of identical copies of the apparatus. In our case the two averages are different in principle, although they coincide for practical purposes, unless  experiments are performed in time intervals shorter than $\Delta t$.
We continue nevertheless, to use the double averaging notation, to underline the conceptual difference.

Suppose we prepare the system at initial time $t=0$. If $\hbar$ fluctuates, position measurements receive additional contributions when averaging over $\varepsilon$ distribution. For mean position value this term is  $\overline{\sqrt{1+\varepsilon}} \sim 1+ {\cal O} (\sigma)$, which is time independent, since fluctuation averages  are time translation invariant. The same results holds at some final time $t$, again by time translation invariance. This implies that measurements of the mean distance travelled by a particle is insensitive to this order $\sigma$ factor, which cancels in the difference $\overline{\langle {\bf x}\rangle}_\psi(t)-\overline{\langle {\bf x}\rangle}_\psi(0)$. The observable effect of averaging over $\varepsilon$ is only contained in the $\varepsilon$ term in the state time evolution, see equation (\ref{wf}). Same holds for position uncertainty and momentum measurements.

We choose for reference a gaussian profile peaked at $\overline{p}$ and variance $\delta^2$
\be
\psi(x,t) = \int \, \frac{d p_0}{\sqrt{2 \pi}} \frac{1}{( \pi \delta^2)^{1/4}} e^{-\frac{(p_0-\overline{p})^2}{2 \delta^2}  + i p_0 x/\hbar -i p_0^2\left( t +  \int_0^t \varepsilon(t') \, dt' \right)/(2 m \hbar)} \, . \label{psit}
\ee
Using equations (\ref{mean0},\ref{epsprop}), the mean distance travelled by a particle and uncertainty read
\bea
\overline{\langle {\bf x}\rangle}_\psi(t)-\overline{\langle {\bf x}\rangle}_\psi(0)&=&\frac{\overline{p}}{m} t\, ,\\
(\Delta{\bf x})^2_\psi(t)-(\Delta{\bf x})^2_\psi(0)&=&\frac{\delta^2}{2 m^2} t^2 + \frac{\overline{p}^2+\delta^2/2}{m^2} \tau\, t\, . \label{delta2massive}
\eea
Squared uncertainty displays a new contribution, linear in time. This is much alike a Brownian motion with diffusion coefficient 
\be
D=\frac{\overline{p}^2+\delta^2/2}{2 m^2} \tau\, .
\ee
For $\delta\ll\overline{p}$, $D$ can be regarded as due to {\it scatterings} with mean free path  $(\overline{p}/m) \tau$.

First term in r.h.s.\ of equation (\ref{delta2massive}) is typically expected to be dominant over the genuine new effect, but considering massless particles, as photons, the linear dispersion relation leads to no ${\cal O}(t^2)$ wave packet spread. A gaussian wave packet reads in this case
\be
\psi(x,t) = \int \, \frac{d p_0}{\sqrt{2 \pi}} \frac{1}{( \pi \delta^2)^{1/4}} e^{-\frac{(p_0-\overline{p})^2}{2 \delta^2}  + i p_0 x/\hbar -i c p_0 \int_0^t \sqrt{1+\varepsilon(t')} dt' /\hbar} \, . \label{psitphoton}
\ee
and to leading (non trivial) order in the expansion of $\sqrt{1+\varepsilon(t)}$ one finds
\bea
\overline{\langle {\bf x}\rangle}_\psi(t)-\overline{\langle {\bf x}\rangle}_\psi(0)&=&c\, t\, ,\\
(\Delta{\bf x})^2_\psi(t)-(\Delta{\bf x})^2_\psi(0)&=&\frac{c^2 \,\tau \,t}4\, . \label{delta2massless}
\eea

This {\it random walk} behavior leads to interesting effects in interference experiments. Consider a light wave packet impinging a plate pierced by two slits, and then observed on a screen, producing an interference pattern. Since waves are detected at some fixed distance $L$ from the plate, the $\hbar$ stochastic nature can be effectively viewed as a change $\delta t$ of the travel time, with variance, from equation (\ref{delta2massless}) 
\be
\overline{\delta t^2} = \frac{\tau \, t}4 = \frac{\tau\, L}{4c} \, , \label{deltat}
\ee 
with $t=L/c$ the time mean value, neglecting the distance between the slits compared to $L$. Assuming a plane wave with frequency $\omega$, light intensity $I$ say, at the point on the screen equidistant from the two slits is
\be
I \propto \frac{1}{4} \left| e^{- i \omega (t +\delta t_1)} + e^{- i \omega (t +\delta t_2)} \right|^2 = \frac12 \left(1+ \cos \left[ \omega ( \delta t_1 -\delta t_2) \right]\right) \, ,\label{interference}
\ee
where the two contributions come from the two slits and $\delta t_{1,2}$ are their (uncorrelated) time shift along the path from slits to the screen. In the standard case the two waves show a constructive interference. Here, averaging over $\delta t_{1,2}$, from equation (\ref{deltat})
\be
I \propto \frac12 \left(1+ \exp \left(-\frac{\omega^2 \tau L}{4c} \right) \right) \, . \label{power}
\ee
The interference term decays exponentially for large $L$, and asymptotically intensity behaves as the two waves {\it were not interfering}. This is a genuine effect of Planck constant fuzziness, which destroys wave coherence on distances $\omega^2 \tau L/(4c) \geq1$. Reasoning in a similar way one finds that the whole interference pattern changes. 

Notice that, using equation (\ref{delta2massless}), the effect can be cast in terms of a fluctuating light speed with variance $(\Delta c)^2$ 
\be
\frac{(\Delta c)^2}{c^2} = \frac{c \, \tau}{4L} \, ,
\ee
Of course, this holds for relativistic particles only, but not in general.

The result of equation (\ref{power}) could be tested in long--baseline interferometric experiments, like Virgo or Ligo \cite{Virgo,Ligo}, aimed at detecting gravitational waves from astrophysical sources. To this end, all noises, mechanical, seismic etc., are kept under an  exquisite control, sensitivity being eventually limited in the relevant frequency range by the irreducible shot noise contribution, due to photon number $N$ (or wave packet phase) Poisson fluctuations in light bunches, $\Delta N = \sqrt{N}$. For a small $\tau$, equation (\ref{power}) leads to 
\begin{equation}
\frac{\Delta I}I \simeq -\omega^2  \frac{\tau}4 \frac{L}{c} \, , \label{effect}
\end{equation}
which assuming no detection of $\tau$ related effects, should be smaller than shot noise (sn) 
\begin{equation}
\omega^2  \frac\tau 4 \frac{L}{c}  < \left(\frac{\Delta I}{I}\right)_{{\rm sn}}= \frac{\Delta N}N = \frac{1}{\sqrt{N}}= \sqrt{\frac{h \nu \Delta \nu}{I}} \, , \label{sn}
\end{equation} 
with $\nu$ and $\Delta \nu$ the light frequency  and bunch bandwidth, or
\begin{equation} 
\tau < 4 \cdot 10^{-34} \, \frac{\mbox{km}}{L} \left(\frac{10^{14} \mbox{Hz}}{\nu}\right)^{3/2} \sqrt{\frac{\Delta \nu}{\mbox{Hz}}}\sqrt{\frac{\mbox{10 W}}{I}} \, .
\end{equation}
We normalized to visible light frequency and status of the art values for $\Delta \nu$, $L$ and power $I$. 
This bound would translate into a lower limit for an effective energy scale of $\hbar$ dynamics $\Lambda= \hbar/\tau\gtrsim 10^{10}$ GeV. 

\subsection{Harmonic oscillator and coherent light in cavities}
\label{ho}
For a one--dimensional harmonic oscillator with frequency $\omega$ and mass $m$ the Hamiltonian reads
\be
\mbox{{\bf H}} = \frac{1}{2 m}(1 + \varepsilon) {\bf p}^2_0 + \frac{m \omega^2}{2} (1+ \varepsilon) {\bf x}^2_0\, ,
\label{hoscill}
\ee
i.e. a standard harmonic oscillator with time dependent mass and frequency, $M=m/(1+\varepsilon)$ and $\Omega=\omega(1+\varepsilon)$,~\cite{caldirola,kanai}, with $M \Omega= m \omega$ a constant. The Hamiltonian depends on time via an overall multiplicative factor, and the commutator $\left[\mbox{{\bf H}}(t),\mbox{{\bf H}}(t') \right]$ vanishes, implying that Dyson series for time evolution operator can be computed explicitly
\be 
U(t) = \exp \left( - \frac{i}{\hbar} \int_0^t H(t') dt' \right) \, . 
\ee
Defining, with standard normalization, creation operator 
\be
{\bf a} = \sqrt{\frac{m \omega}{2 \hbar}} {\bf x}_0 + i \frac{1}{\sqrt{2 m \hbar \omega}}{\bf p}_0 \,,
\label{a}
\ee
we have
\be \mbox{{\bf H}} = \hbar \omega (1 + \varepsilon) \left( {\bf a}^\dagger {\bf a} + \frac12 \right) \, .
\ee
Notice that ${\bf a}$ and ${\bf a}^\dagger$ do not explicitly depend on time and obey
\be
\left[ {\bf a}, {\bf a}^\dagger \right] = 1\, .
\label{comma}
\ee
The ${\bf a}$ equation of motion 
\be
\frac{d{\bf a}(t)}{dt} = -i \omega \,(1 +\epsilon(t))\, {\bf  a}(t) \, ,
\label{eom}
\ee
has the formal solution 
\be
{\bf a}(t) = {\bf a}(0) e^{ -i \, \omega t} \sum_n \frac{(-i \omega)^n}{n!} \int_0^t dt_1 .... \int_0^t  dt_n \epsilon(t_1)....\epsilon(t_n) \, . \label{formal}
\ee
Averaging over  $\epsilon(t)$ probability distribution and computing $n$--point correlation functions in terms of two--point correlation {\it \`{a} la} Wick
\bea
 \overline{{\bf a}(t)} &=& {\bf a}(0) e^{ -i \, \omega t} \sum_k \frac{(-\omega^2)^k}{2k!} (2k-1)!! \,(\tau\, t)^k   \nonumber \\
&=& {\bf a}(0) e^{ -i \, \omega t} \sum_k \frac{(-\omega^2 \tau t)^k}{ 2^k k!} = {\bf a}(0) e^{ -i \, \omega t} e^{-\omega^2 \tau t/2}\, .
\label{resa}
\eea 
Apart from standard oscillatory term, evolution is exponentially damped on time--scales larger than the characteristic time $2 (\omega^2 \tau)^{-1}$. 

Consider now a coherent state $|\lambda \rangle$ at time $t=0$. With no loss of generality we take $\lambda$ real. As discussed already, position/momentum measurements at some particular time amounts to measure ${\bf x}_0$ and  ${\bf p}_0$. From  equation (\ref{formal}) after averaging over $\varepsilon$ distribution 
\bea 
 \overline{\langle  \bf{x} \rangle}_\lambda (t)&=&\sqrt{\frac{2 \hbar}{m \omega}} \,\lambda \,\cos ( \omega t) e^{-\omega^2 \tau t/2} \, ,  \nonumber \\
 \overline{\langle  \bf{p} \rangle}_\lambda (t) &=&-\sqrt{2 \hbar m \omega}\, \lambda \, \sin ( \omega t) e^{-\omega^2 \tau t/2}\, , \nonumber \\
 \overline{\langle  \bf{x}^2 \rangle}_\lambda (t)&=&\frac{ \hbar}{m \omega} \left[ \frac12 +\lambda^2 \left(1+ \cos( 2 \omega t) e^{-\omega^2 \tau t} \right) \right] \, ,\nonumber \\
 \overline{\langle  \bf{p}^2 \rangle}_\lambda (t)&=&m \hbar \omega \left[ \frac12 +\lambda^2 \left(1- \cos( 2 \omega t) e^{-\omega^2 \tau t} \right) \right] \, , \nonumber \\
\label{various}
\eea
so that coherent states do not saturate the lower limit of Heisenberg relation as time flows, unless $\lambda=0$, 
\be
\overline{\Delta \bf{x}}_\lambda \overline{\Delta \bf{p}}_\lambda = \frac\hbar2 \left[1+ 2 \lambda^2\,\left(1-e^{-\omega^2 \tau t} \right) \right].
\label{uncert} 
\ee
This effect is in all similar to decoherence processes affecting the oscillator phase  but leaving its energy unperturbed.
We show position mean value, squared mean value and uncertainty in Figure 1. 
The growing behaviour of $\overline{\Delta {\bf x}^2}_\lambda$  can be appreciated provided the state remains in a coherent configuration for  times $t \omega \geq (\tau \omega)^{-1}$.  

These features could be constrained using optical cavities. Excited by an external coherent beam, they can store coherence properties of electromagnetic field for long times, and consist in spatial confinement between two highly reflecting surfaces of a well defined propagation mode. Commonly used in laser/maser physics and in optical experiments where high spectral spatial purities are required (high resolution spectroscopy, interferometry, quantum optics, etc.), their confinement ability can be quantified, as for mechanical oscillators, in terms of quality factor $Q=\omega t_{c}$, the ratio between energy lost in a cycle to the energy stored in the cavity, with $\omega$ the  frequency and $t_{c}$ the cavity decay time. Using supermirrors, at optical frequency, values of $Q \sim 10^{15}$ are accessible \cite{Young1999}.

A single mode coherent state of an electromagnetic harmonic oscillator is a  minimum uncertainty states for any pair of orthogonal field quadrature operators \cite{Sudarshan1963,Glauber1963}, the analogue of position and momentum operators. In phase space this state is represented by a two dimensional Gaussian distribution with equal variances at all direction.
To keep a stricter analogy with the harmonic oscillator discussed here, we define field quadratures as ${\bf X}=\sqrt{\hbar/2}\left({\bf a}+{\bf a}^{\dagger}\right)$ and ${\bf Y}=i\sqrt{\hbar/2}\left({\bf a}^{\dagger}-{\bf a}\right)$. Measurements of the uncertainty for a given quadrature of an electromagnetic  mode can be obtained by a homodyne detector.

For a random $\hbar$, coherent configuration of radiation do not saturate the lower bound $\hbar/2$ for $\Delta {\bf X} \,\Delta {\bf Y}$, which monotonically increases and is related to $|\lambda|^2$, the mean photon occupation number, see equation (\ref{uncert}). Measurements in resonant cavities however, are limited by $t_c$. For $t>t_c$ the coherent electromagnetic field escapes from cavity due to unavoidable couplings to the external thermal bath, and the system evolves towards vacuum state $\lambda=0$. 
\begin{figure}
\begin{center}
\includegraphics[width=.8\textwidth]{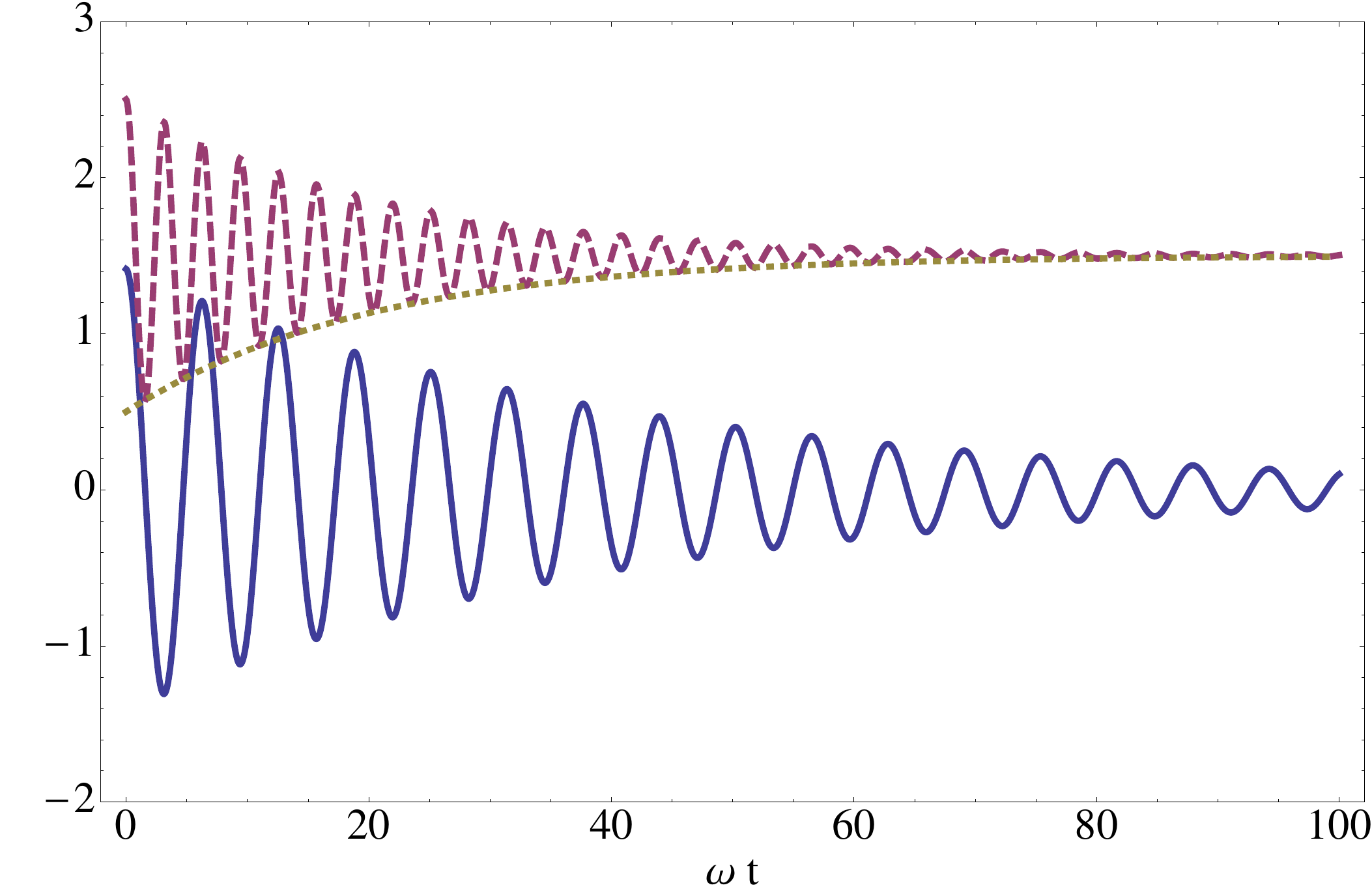}
\caption{\label{fig1} \sl The time evolution of position  mean value $\overline{\langle  \bf{x} \rangle}_\lambda$  (solid), $\overline{\langle  \bf{x}^2 \rangle}_\lambda$ (long-dashed) and squared uncertainty $\overline{\Delta \bf{x}^2}_\lambda$(short-dashed) for a coherent state with $\lambda=1$. Values are in units of appropriate powers of the length unit $\sqrt{\hbar/(m \omega)}$. We have chosen an unrealistic large value $\omega \tau=0.05$ to emphasize the non standard time behavior with respect to the case of a constant $\hbar$. }
\end{center}
\end{figure}
To account for this effect we modify equation (\ref{uncert}) by introducing an exponentially damped $\lambda$
\be
\overline{\Delta \bf{X}}_\lambda \overline{\Delta \bf{Y}}_\lambda = \frac\hbar2 \left[1+ 2 \lambda^2 e^{-2 t/t_c}\,\left(1-e^{-\omega^2 \tau t} \right) \right].
\label{uncert2} 
\ee
This approximation is satisfied if $t_c=Q/\omega$ is much larger than $\tau$, so that $\lambda$ adiabatically decays on  $\tau$ time scales. The r.h.s.\ of equation (\ref{uncert2}) grows for small times, reaches a maximum at $t_*$
\be
\left. \overline{\Delta \bf{X}}_\lambda \overline{\Delta \bf{Y}}_\lambda \right|_{t_*} = \frac\hbar2 \left[1+  \lambda^2 Q \omega \tau \left(\frac{2}{2+Q \omega \tau} \right)^{(2+ Q \omega \tau)/(Q \omega \tau)} \right] \, , \label{max}
\ee
and eventually decays towards the standard value $\hbar/2$. For $Q \omega \tau\ll1$, $t_* \simeq t_c/2-Q^2 \,\tau/8$.

Consider  one measures the uncertainty product with some error. In the standard approach, this product for a coherent state gives  $\hbar/2$, so is a way to determine Planck constant, with some uncertainty, $\hbar \pm \Delta \hbar$. If the time behaviour of equation (\ref{uncert2}) is undetected, in particular the peak at $t_*$, this means that $\tau$ should be sufficiently small
\be
\lambda^2 Q \omega \tau \left(\frac{2}{2+Q \omega \tau} \right)^{(2+ Q \omega \tau)/(Q \omega \tau)}  < \frac{\Delta \hbar} \hbar \, , \label{bound}
\ee
To have an order of magnitude of this bound, we take $\lambda = 1$. Choosing $Q = 10^{15}$, $\omega \approx 3\cdot 10^{15}$ Hz, \cite{Young1999}, and a measurement uncertainty  $\Delta \hbar/\hbar \sim 1 \% $ we obtain
\be 
\tau < 10^{-32} \, \mbox{s} \, ,
\ee
or, in terms of energy scale, $\Lambda= \hbar/\tau > 10^8$ GeV. For smaller $\Delta \hbar$, the bounds on $\tau$ scales approximately as $(\Delta \hbar /\hbar) /(Q \omega)$, see equation (\ref{bound}).

\section{Concluding remarks}

We have considered the possibility that Planck constant is a randomly fluctuating quantity. These fluctuations are viewed as the manifestation at low energies of some fuzzy structure of spacetime and phase space at very high energy scales $E > \Lambda$, and are found to change the standard results of quantum mechanics, such as the Heisenberg uncertainty relation and interference of massive or massless particle beams. The novel effects can be effectively regarded as due to the coupling of a particular system, such as an harmonic oscillator or a free particle, with some dynamical field. Even if this field has reached a stable background configuration, we expect its value to undergo small quantum fluctuations, which introduce a noise in the time evolution of the system. In particular, we have studied the case in which this noise shows up in terms of a random Planck constant with a variance times typical correlation time parameter $\tau=\sigma^2 \Delta t$.
Interestingly, the model can be strongly constrained by future experiments with long baseline interferometers or on coherent light cavities, and adopting present status of the art parameters for these kind of experiments, we found that $\tau$ could be bound in dedicated measurements to a very tiny value, of order of $10^{-34}-10^{-32}$ s. These time values translate into energy scales $\Lambda \sim \hbar/\tau$ larger than $10^8-10^{10}$ GeV, which are well below the Planck mass scale, $10^{19}$ GeV, yet above the energy range which can be explored with present accelerators, like LHC at CERN, and presumably also unaccessible to  next generation accelerator programs. 

Of course, since Planck constant is ubiquitous, there is potentially a very large number of different dynamical systems and experimental approaches which may be exploited, and which were not covered here. Possibly, some of them may be even more powerful than those we discussed in constraining a stochastic $\hbar$. To study them all would entail the reconsideration of the basics of all quantum phenomena, an enterprise which is well beyond the scope of this paper.

\subsubsection*{Acknowledgments}
F.L.\ and G.M.\ are supported by INFN, I.S.'s GEOSYMQFT and  TASP, respectively. This article is based upon work from COST Action MP1405 QSPACE, supported by COST (European Cooperation in Science and Technology). F.L.\ is partially supported by CUR Generalitat de Catalunya under projects FPA2013-46570 and 2014~SGR~104, MDM-2014-0369 of ICCUB (Unidad de Excelencia `Maria de Maeztu'), and by UniNA and Compagnia di San Paolo under the grant Programma STAR 2013.

\end{document}